\documentclass[12pt]{iopart}
\usepackage{iopams}
\usepackage{epsfig}
\usepackage{graphics}

\begin{document}

\title[Coherence of qubits based on single Ca$^+$ ions]
{Coherence of qubits based on single Ca$^+$ ions}

\author{F.~Schmidt-Kaler\footnote[3]{To whom correspondence should be
addressed (ferdinand.schmidt-kaler@uibk.ac.at)}, S.~Gulde,
M.~Riebe, T.~Deuschle, A.~Kreuter, G.~Lancaster, C.~Becher,
J.~Eschner, H.~H\"affner, and R.~Blatt}

\address{Institut f{\"ur} Experimentalphysik, 6020 Innsbruck,
Austria}

\begin{abstract}
Two-level ionic systems, where quantum information is encoded in
long lived states (qubits), are discussed extensively for quantum
information processing. We present a collection of measurements
which characterize the stability of a qubit based on the
$S_{1/2}$--$D_{5/2}$ transition of single $^{40}$Ca$^+$ ions in a
linear Paul trap. We find coherence times of $\simeq$1~ms, discuss
the main technical limitations and outline possible improvements.
\end{abstract}



\section{Introduction}
The concept of quantum computing is based on the coherent
manipulation of quantum bits (qubits), which carry the information
in a superposition of two quantum mechanical states $\{|0\rangle$,
$|1\rangle\}$ \cite{PhysQI00,CHUANG00,VINCENCO01}. A quantum
register is composed out of $N$ qubits. Prior to the computation,
the state of all qubits is set to a well defined initial value.
During the following computation process, a large number of
quantum gate operations is performed on these qubits. This
sequence of logic gate operations is determined by the specific
task, following the desired quantum algorithm. Ideally, during the
course of this algorithm the quantum state of the system follows a
fully unitarian, thus time-reversible path in the
$2^N$-dimensional Hilbert space, free of any decoherence. Finally,
the qubits are projected in the computational basis and the
outcome of the algorithm is measured.

However, different from this ideal situation, decoherence will
occur and cause errors during the computation process. Thus,
highly-entangled quantum states generated during the computation
are destroyed and the operation of the quantum computer is
affected \cite{HAROCHE96,MERMIN01}. Any coupling of the quantum
computer to the environment causes decoherence and it seems an
impossible task to construct a quantum computer working in an
entirely coherent way. Is quantum computing impossible? In order
to overcome the problem of decoherence, quantum-error-correction
schemes have been proposed \cite{SHOR95,STEANE96} which lift the
constraint on coherence to an acceptable error rate of about 1 in
10$^5$ computational steps \cite{PRESKILL98,CHUANG00}. Under this
condition, error correction schemes predict the stabilization of
any quantum computation process.

Presently, a number of proposals are discussed for a future
realization of quantum computation, emanating from various fields
of physics. We expect that a future quantum computer has to
fulfill the following list of requirements \cite{VINCENCO01} and
any proposal for a future quantum computer should address them.
Each system which is proposed for an implementation of a quantum
computer will have to provide

\begin{itemize}
  \item a scalable physical system with well characterized qubits,
  \item the ability to initialize the state of the qubits,
  \item a coherence time much longer than operation time,
  \item an universal set of quantum gates:
        single bit and two bit gates,
  \item a qubit-specific measurement.
\end{itemize}

In this paper we discuss strings of ions for quantum computation
\cite{SASURA02}: Ion strings can be stored in linear Paul traps
such that they arrange in string, representing a quantum register
\cite{PhysQI00}. They can be optically pumped and optically
cooled, such that the register is initialized. Ions are kept under
ultra-high vaccum conditions, thus we expect a long coherence
time. Initiated by the proposal by Cirac and Zoller
\cite{CIRAC95}, other procedures for quantum gate operations have
been developed \cite{SOERENSEN99} and realized \cite{SACKETT00}.
Single qubit operations performed by Rabi oszillations have been
shown as well \cite{NAEGERL98a,ROHDE01}. For the qubit specific
measurement, the 100\% efficient electron shelving technique is
employed \cite{shelving,ROOS99}. Thus, ion trap quantum computing
has already left the status of a theoretical concept in so far as
experiments with a small number of qubits have been perfomed and
the properties of the system have been studied in some detail
\cite{ROHDE01,ROOS99,TURCHETTE00}.

Here, we focus the discussion to the third item of the above list
of requirements: What are the sources of decoherence and how can
we investigate them quantitatively? Under the conditions set by
the actual experiments \cite{WEBBOULDER,WEBIBK}, we find that the
time required for a single logic operation is on the order of
$10^1 - 10^2~\mu$s \cite{STEANE00}. Thus, if we demand that the
error probability due to decoherence is smaller than 1 in 10$^5$
we find that a coherence time of at least 1..10~s is required for
a successful application of quantum error correction. In this
paper we try to give a deeper insight into the present, mostly
technical, limitations of the qubit coherence.

Typically, with trapped ions, a qubit is encoded in atomic
transitions involving levels with extremely low radiative decay:
Hyperfine ground states are used as qubit-bases and manipulated
via a far off-resonant Raman transition \cite{WINE98}. Optionally,
the qubit is encoded in a superposition of a ground state
$|\textrm{S}\rangle$ and a long lived metastable state
$|\textrm{D}\rangle$ \cite{ROOS99,NAEGERL00}, and manipulated on
an optical quadrupole transition. For two-bit gate operations
\cite{CIRAC95,SOERENSEN99}, the excitation of motional quantum
states \cite{JAMES98} in a string of ions \cite{NAEGERL98b} is
used.

In this paper we consider the qubit transition in $^{40}$Ca$^+$,
from the S$_{1/2}$ ground state to the metastable D$_{5/2}$ state
(lifetime~ $\simeq$~$1s$), and recent experimental investigations
of the Innsbruck group \cite{WEBIBK}. The paper is organized as
follows: The first section contains a brief description of the
experimental set-up and the experimental techniques which are used
for trapping, cooling and observing ion strings. In the second
section we discuss the coherence of the internal, electronic qubit
state $\{|S\rangle,|D\rangle\}$, and  in a third section we
investigate coherence properties of the lowest motional states
$\{|0\rangle,|1\rangle\}$. The final section sketches a proposal
which possibly allows more than $10^{2}$ quantum logic operation
and coherence times exceeding 1~s, using trapped $^{43}$Ca$^+$
ions.

\section{Experimental setup}
\begin{figure}[t]
\begin{minipage}{0.45\linewidth}
\begin{center}
\epsfig{file=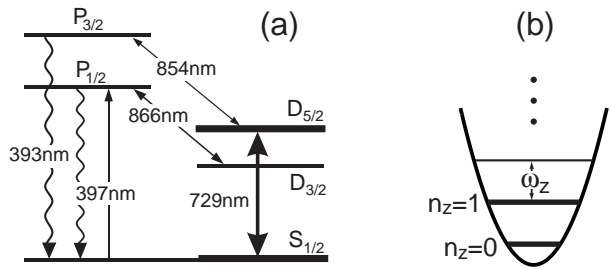,width=8cm}
\end{center}
\end{minipage}
\begin{minipage}{0.55\linewidth}
\caption{\label{ca} (a) Level scheme of Ca$^+$ ion. Superpositions
of S$_{1/2}$ and D$_{5/2}$ serve as qubits. (b) Vibrational states
in the harmonic trapping potential, Fock states $|n=0\rangle$ and
$|n=1\rangle$ serve as qubits.}
\end{minipage}
\end{figure}

For the experiments, a single $^{40}$Ca$^+$ ion is, or a string of
a few ions are, stored in a linear Paul trap. The trapped
$^{40}$Ca$^+$ ion has a single valence electron and no hyperfine
structure (see Fig.~\ref{ca}a). We perform Doppler-cooling on the
$\textrm{S}_{1/2} - \textrm{P}_{1/2}$ transition at 397~nm. Diode
lasers at 866~nm and 854~nm prevent optical pumping into the D
states. For sideband cooling and for quantum information
processing \cite{ROOS99}, we excite the S$_{1/2}$ to D$_{5/2}$
transition with a Ti:Sapphire laser near 729~nm (linewidth $\leq
100$~Hz). A constant magnetic field of 2.4~G splits the 10 Zeeman
components of the S$_{1/2}\leftrightarrow$ D$_{5/2}$ multiplet.
Depending on the chosen geometry and polarization, the excitation
of $\Delta m = 0, \pm1$ and $\pm~2$ transitions is allowed. We
detect whether a transition to D$_{5/2}$ has occurred by applying
the laser beams at 397~nm and 866~nm and monitoring the
fluorescence of the ion on a photomultiplier (electron shelving
technique \cite{shelving}). The internal state of the ion is
discriminated with an efficiency close to 100$\%$ within 3~ms
\cite{ROOS99}. The Paul trap is built with four blades separated
by 2~mm for radial confinement and two tips separated by 5~mm for
axial confinement. Under typical operating conditions we observe
axial and radial motional frequencies $(\omega_{ax},
\omega_r)/2\pi =$~(1.7, 5.0)~MHz.

\subsection{Measurement cycle \label{cycle}}
The measurement cycle (total duration 20~ms) consists of four
consecutive steps: (I) Doppler cooling (laser 397~nm, 866~nm and
854~nm on) leads to low thermal vibrational states of axial and
radial modes with $\langle n_{ax} \rangle \approx 15$ and $\langle
n_{r} \rangle \approx 3$ phonons. (II) Sideband cooling of the
axial motion is performed on the $|\textrm{S}_{1/2},
m=-\frac{1}{2} \rangle \leftrightarrow |\textrm{D}_{5/2},
m'=-\frac{5}{2} \rangle$ transition, leading to more than 99~$\%$
ground state population. Pumping into $|\textrm{S}_{1/2},
m=+\frac{1}{2} \rangle$ is counteracted by several short pulses of
$\sigma^-$ radiation at 397~nm. (III) Manipulation of the qubit
state $|\{S,D\},\{n=0,1\} \rangle$ with radiation near 729~nm.
(IV) Final state analysis: The ion's fluorescence is collected
under excitation with laser light at 397~nm and 866~nm and thus
the internal state is detected. This sequence is repeated
100~times to measure the D$_{5/2}$ state excitation probability
$P_D$. All experiments, described in the following, are performed
using this sequence. To investigate different sources for and
sorts of decoherence, we only modify step (III) of the sequence
according to the specific task.

\section{Coherence of the internal qubit state}

Qubits are represented by the electronic state $\alpha |S\rangle +
\beta |D\rangle$ of each ion in the linear string where $\alpha^2
+ \beta^2 =1$. Decoherence leads to the decay of the quantum
mechanical phase relations transforming the above state into an
incoherent mixture. Various possible reasons for decoherence on
the qubit transition S$_{1/2}$ -- D$_{5/2}$ can be expected:

\begin{itemize}
  \item Qubit energy levels fluctuate via the Zeeman effect
  caused by ambient  magnetic field fluctuations in the ion trap,
  \item the laser light driving the qubit transition
  fluctuates in frequency and light intensity,
  \item the upper qubit basis state D$_{5/2}$ decays spontaneously (1~s
  lifetime).
\end{itemize}

While the first items are due to technical shortcomings,  only the
third item is a physical limit for the coherence time. In the
following sections, we present a variety of experiments which
investigate (and discriminate between) different sources of
decoherence as listed above.

\subsection{Noise components at 50~Hz \label{50Hz}}

To test the influence of the frequency noise components at
$50$~Hz, e.g. caused by ambient magnetic field fluctuations in the
ion trap, we trigger the experimental sequence synchronized to the
$50$~Hz frequency from the power line. If the 50~Hz noise
components are dominant, the laser will excite the qubit
transition at an instant of time when the ion is exposed to about
the same ambient magnetic field. The magnitude of the magnetic
field variation is examined by shifting the excitation pulse in
time with respect to the 50~Hz line trigger and measuring the
resonance frequency of the S$_{1/2}(m=-1/2)$ -- D$_{5/2}(m'=-5/2)$
with a laser detuning near $\Delta$=0 ("carrier transition"). The
linear Zeeman shift depends on the magnetic $g$-factors for both
states which are involved, $g_{S1/2}$=2 and $g_{D5/2}$=6/5, and
the magnetic  quantum numbers $m$ for initial and final state.
Therefore, Zeeman shifts are most sensitively measured via the $m=
-1/2 \rightarrow -5/2$ Zeeman component ($\Delta m$=2) where a
field of 1~mGauss corresponds to 4.2~kHz shift of the resonance
center.

\begin{figure}[t,b]
\begin{minipage}{0.45\linewidth}
\begin{center}
\epsfig{file=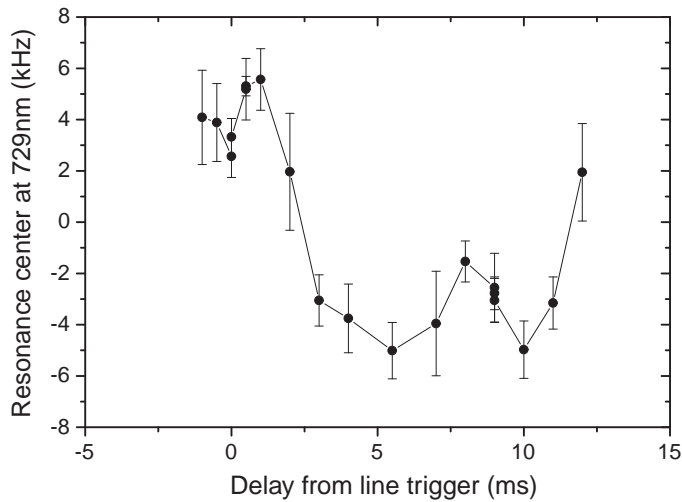, height=7cm}
\end{center}
\end{minipage}
\begin{minipage}{0.55\linewidth}
\caption{\label{linetrigger} The magnetic field fluctuations at 50~Hz shift
the $S_{1/2}, m=-1/2 \rightarrow D_{5/2}, m=-5/2$ carrier resonance, as the
delay between the laser pulse and the line trigger is varied. The interaction
time is $\tau$=1~ms. Note that the vertical bars indicate the width of the
observed resonance, not the error of its center frequency. }
\end{minipage}
\end{figure}

For this, the laser frequency is varied over the resonance of the
S$_{1/2}(m=-1/2)$ -- D$_{5/2}(m'=-5/2)$ resonance. We record the
excitation probability to the upper state, after a 1~ms laser
pulse (step III of the experimental sequence). The line center
position is determined from a fit to the data. We thus observe,
that the center of the resonance line fluctuates within a
bandwidth $\simeq \pm$~5~kHz (Fig.~\ref{linetrigger}). Even though
the details of these fluctuations are not reproduced from day to
day, its amplitude remains fairly constant with approximately one
mGauss.

To avoid the qubits's dephasing due to 50~Hz noise, {\em all}
experiments described below are triggered to the power line
frequency. Secondly, to avoid the disturbing influence of the
linear Zeeman effect as much as possible, we use the
S$_{1/2}(m=-1/2)$ -- D$_{5/2}(m'=-1/2)$ transition ($\Delta m$=0)
for quantum gate operations, with a five times smaller
susceptibility for magnetic field fluctuations.

\subsection{Rabi oscillations}

\begin{figure}[t,b]
\begin{minipage}{0.45\linewidth}
\begin{center}
\epsfig{file=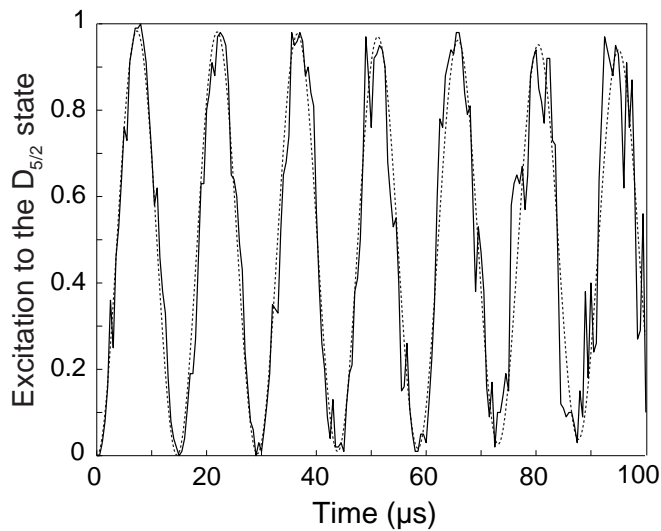, height=7cm}
\end{center}
\end{minipage}
\begin{minipage}{0.55\linewidth}
\caption{\label{Rabi} Rabi oscillations on the carrier ($\Delta=0$) of the
S$_{1/2}(m=-1/2)$ to D$_{5/2}(m=-1/2)$ transition. The numerical simulation
(dashed) takes into account 3\% laser intensity fluctuations and phonon
numbers after Doppler and sideband cooling of $n_{rad}=7, n_{ax}=0$. From the
geometry of the trap and the trap frequencies we calculate the Lamb-Dicke
factors for the excitation $\eta_{ax}=0.068$ and $\eta_{rad}=0.016$.}
\end{minipage}
\end{figure}

The most simple manipulation consists of a single laser pulse in
step~(III). While the laser detuning $\Delta=0$ is kept fixed, the
pulse length $\tau$ is varied. The excitation probability to the
D$_{5/2}$ state is plotted versus the duration $\tau$, and we
observe Rabi oscillations as shown in Fig.~\ref{Rabi}. This way
single bit rotations are realized. From the measured contrast of
the oscillations we derermine the quality of qubit rotations. For
the given laser intensity here, a $\pi$ rotation is achieved after
about 7~$\mu$s, to implement the logic NOT operation. Even for a
$10\pi$ rotation we observe a contrast of better than 94\% at
times near 75~$\mu$s. We identify two reasons which limit the
contrast: i)~Laser intensity fluctuations cause slightly different
Rabi frequencies from shot to shot. As each data point in
Fig.~\ref{Rabi} is taken as the average over 100 experimental
realizations of the sequence (sect.~\ref{cycle}) the contrast is
reduced. ii)~A second limitation of the quality of single qubit
operations is caused by residual small phonon numbers in thermally
occupied vibrational modes. Other than the vibrational mode used
for the quantum gate operations, typically not all modes are
cooled to the vibrational ground state $|n=0\rangle$. As the Rabi
frequency on the carrier ($\Delta$=0) transition depends weakly on
the phonon occupation number in all those modes ("spectator
modes"), averaging over their thermal distributions leads to a
reduced contrast \cite{WINE98}. However, this problem can be
solved using newly developed cooling techniques \cite{MORIGI00},
which have been demonstrated recently \cite{ROOS00}. It can be
shown that these cooling techniques will reduce the thermal
occupation of all spectator modes of an ion string well below one
\cite{SCHMIDT01}, where their effect on the contrast of Rabi
oscillations becomes negligible.

\subsection{Ramsey spectroscopy on the S$_{1/2}$ to D$_{5/2}$ transition}

Ramsey spectroscopy is perfectly suited for a test of the qubit's
decoherence. A first $\pi/2$ pulse on the carrier ($\Delta \simeq
0$) excites the state $|S\rangle$ and $|D\rangle$ to a coherent
superposition. After a second $\pi/2$ pulse, applied after a
waiting time of $t$, the resulting state is projected into the
basis $\{|S\rangle,|D\rangle\}$ by the final state analysis.
Varying the laser frequency detuning $\Delta$ yields a Ramsey
pattern. Ideally, the excitation $P_D(\Delta)$ to the D$_{5/2}$
state should exhibit a modulation between zero and one when the
detuning of the laser frequency is slightly varied.

\begin{figure}[t,b]
\begin{minipage}{0.6\linewidth}
\begin{center}
\epsfig{file=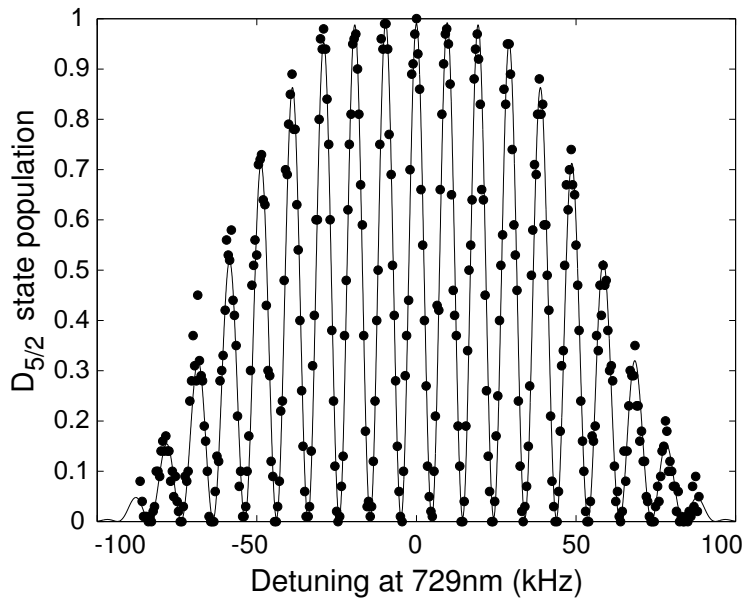, height=8cm}
\end{center}
\end{minipage}
\begin{minipage}{0.4\linewidth}
\caption{\label{Ramsey} Ramsey spectroscopy on the
S$_{1/2}(m=-1/2)$ to D$_{5/2}(m=-1/2)$ carrier transition ($\Delta
\simeq 0$). The scheme consists of a first pulse of 9.5~$\mu$s
duration followed by a second identical pulse after a waiting time
$t$=100~$\mu$s. For the description of the theoretical curve see
the text.}
\end{minipage}
\end{figure}

Experimentally observed Ramsey fringes show $\simeq 99\%$ contrast
(Fig.~\ref{Ramsey}). For the theoretical prediction,
two-level-Bloch equations are solved, keeping the length of the
$\pi/2$-pulses and the waiting time fixed at the experimental
values. Free parameters for the theoretical curve are the pulse
area of the $\pi/2$-pulses of $\Omega_{\pi/2}$= 0.515~$\pi$
(instead of 0.5~$\pi$) and the coherence time, accounting for an
effective laser linewidth of $\nu_{1/2}$=150~Hz.

\begin{figure}[t]
\begin{minipage}{0.45\linewidth}
\begin{center}
\epsfig{file=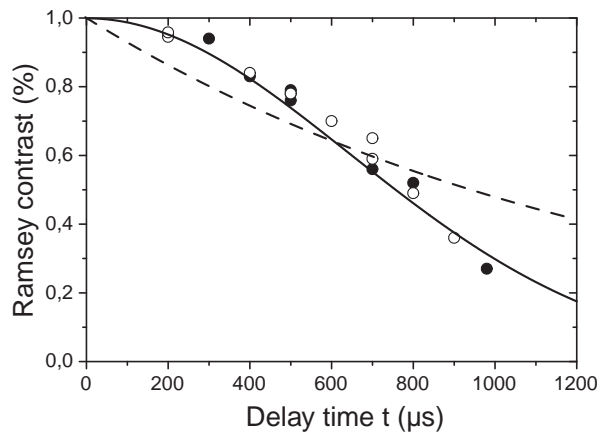, height=6cm}
\end{center}
\end{minipage}
\begin{minipage}{0.55\linewidth}
\caption{\label{RamseyKontr} Contrast of the Ramsey pattern as a function of
waiting time. Solid (open) circles: Data taken without (with) $B$-field
compensation, see text for details. We fit a Gaussian $\sim
\exp^{-(t/\tau)^2}$ with $\tau \simeq$ 0.94(5)~ms. Dashed line: Optimum
exponential fit $\sim \exp^{-t/\tau}$ with the $\tau \simeq$ 1.4(2)~ms.}
\end{minipage}
\end{figure}

If the superposition of the $|S\rangle$ and the $|D\rangle$ states
is exposed to the influence of any decoherence for a long time
duration $t$, causing a dephasing of the qubit levels, this shows
up as a loss of contrast in the observed Ramsey pattern.
Systematically we have varied the waiting time $t$ between both
Ramsey pulses between 100 and 1000~$\mu$s. The observed contrast
$C= (P_D^{max}-P_D^{min})/(P_D^{max}+P_D^{min})$ of the central
fringes is plotted versus the waiting time in
Fig.~\ref{RamseyKontr}. Assuming a white noise model for the
spectral density of frequency fluctuations, one would expect an
exponential $C \sim \exp(-2\pi \nu_{1/2} t)$ to describe the
decrease of contrast, and to yield the Lorentzian linewidth
$\nu_{1/2}$ \cite{SENGSTOCK94}. However, the exponential fit to
our data is very poor, instead we find that a Gaussian, with a
width of $\nu\simeq$ 170(10)~Hz describes our data correctly.
Compared to the Lorentzian noise model, the observed noise shows
an excess of modulation frequencies $\leq$~1.5kHz.

We use an optical fiber to transport the light from the
Ti:Sapphire laser to the ion trap setup. This fibre was identified
as a source of frequency noise with about 20~Hz \cite{DRROHDE01}.
If the fibre induced noise will dominate the frequency noise, we
will apply cancellation techniques as described in
\cite{MA94,YOUNG99}.

\subsection{Active compensation of ambient magnetic field fluctuations}
Since ambient magnetic fluctuations affect the coherence, we have
set up an active magnetic field compensation system \cite{OXFORD}.
This system consists of a flux gate sensor and a control unit
which supplies three orthogonal coils for the cancellation. The
specifications of the compensation device claim a bandwidth of
0.5~Hz to 5~kHz. With the aid of an independent flux gate sensor
we find that the $\simeq$ 1~mGauss ambient field fluctuations
(rms) at 50~Hz are reduced by a factor of $\simeq$~20.

However, no effect of the cancellation system on the Ramsey
contrast is observable as is indicated by Fig.~\ref{RamseyKontr},
by the solid circles. As we measure that the bandwidth of the
cancellation system is sufficient to suppress noise from 50~Hz to
1~kHz, we assume that the remaining sources of magnetic field
fluctuations are localized in the direct vicinity of the trap and
not picked up by the sensor of the cancellation system. In future,
we plan to passively shield the trap from magnetic fluctuations.

\subsection{Raman transitions}

\begin{figure}[t]
\begin{minipage}{0.45\linewidth}
\begin{center}
\epsfig{file=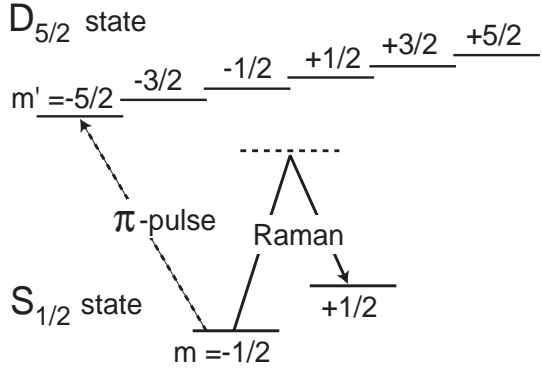, height=5cm}
\end{center}
\end{minipage}
\begin{minipage}{0.55\linewidth}
\caption{\label{raman_schema} Principle of Raman transitions
between Zeeman state S$_{1/2}$, m= -1/2 $\rightarrow$ m= +1/2:
i)~The ion is prepared in the m= -1/2 ground state by optical
pumping. ii)~The Raman pulse is applied to excite the $m= +1/2$
ground state. iii)~A $\pi$-pulse transfers selectively one of the
Zeeman states ($m=-1/2$) to the D$_{5/2}$, m'=5/2 level. After a
successful Raman excitation the ion will be detected in S$_{1/2}$
state, otherwhise step IV of the experimental sequence
(sect.~\ref{cycle}) reveals the D$_{5/2}$ state.}
\end{minipage}
\end{figure}

Further investigations of the magnetic field fluctuations are
performed driving Raman transitions between Zeeman substates
$S_{1/2}: m=-1/2 \rightarrow m=1/2$. For the principle see
Fig.~\ref{raman_schema}.

After the preparation steps~I and II in the experimental sequence,
in step~III we drive the Raman transition. Both laser fields R1
and R2 are generated from the output of the same Ti:Sapphire laser
by means of an acousto-optical modulator (in double pass
configuration), driven with two frequencies. Both beams propagate
in the same optical mode, within the same optical fiber
\cite{MA94,YOUNG99} and illuminate the ion with a resulting
Lamb-Dicke factor of $\eta \simeq$ 0). The Raman detuning is
$\Delta_R$ = 500~kHz. The detuning of R1 is kept fixed, and the
frequency of R2 is varied over the resonance. Both light fields
are switched on and off together. After this Raman pulse, we use a
$\pi$ pulse on the S$_{1/2}, m=-1/2 \rightarrow D_{5/2}, m'=-5/2$
for shelving the $m=-1/2$ Zeeman component. Finally, in step IV,
the excitation probability to the D$_{5/2}$ state is detected.

The obvious advantage is that the Raman excitation technique is
sensitive only to the fluctuations of the ground state Zeeman
levels, but it remains immune to laser frequency fluctuations. Its
second advantage is a potentially higher susceptibility for
magnetic fluctuations, S$_{1/2}: m=-1/2 \rightarrow$ +1/2 with
2.8~kHz/mGauss and D$_{5/2}, m'=-5/2 \rightarrow +3/2$ with
6.72~kHz/mGauss for the transition as compared with the
S$_{1/2}(m=-1/2) \rightarrow $D$_{5/2}(m'=-5/2)$ resonance of
4.2~kHz/mGauss.

We observed Raman spectra and use a 1~ms pulse length for the
excitation. The spectra of show a linewidth of about 2~kHz. As the
width is not given by the Fourier limit, according to the pulse
duration, we conclude that the of magnetic field fluctuations are
important within 1~ms.

\begin{figure}[t,b]
\begin{minipage}{0.45\linewidth}
\begin{center}
\epsfig{file=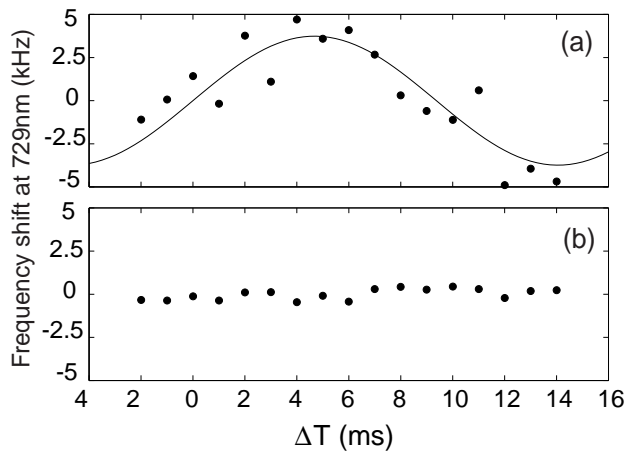, height=6cm}
\end{center}
\end{minipage}
\begin{minipage}{0.55\linewidth}
\caption{\label{B_Komp} Position of the Raman resonance as a
function of laser pulse time delay from the 50~Hz trigger. (a)
Without active magnetic field compensation: We fit a $sine$
function with amplitude 3.8~kHz, corresponding to 0.75(25)~mGauss
field fluctuation. (b) Raman resonance frequency fluctuation using
the active compensation, the standard deviation of the data of
0.3~kHz corresponds to magnetic fluctuations of 0.12~mGauss
residual field fluctuations.}
\end{minipage}
\end{figure}

Caused by the dominant 50~Hz component of the ambient magnetic
field fluctuations, the Raman resonance is shifted, depending on
the delay time from the 50~Hz trigger pulse \ref{50Hz}. We take
Raman spectra for different delays of the excitation time with
respect to the line-trigger, and plot their center frequency
versus the delay time in Fig.~\ref{B_Komp}a. We observe a
modulation with the amplitude of a few kHz at a frequency of 50~Hz
corresponding to a mGauss (rms) magnetic field fluctuation. The
active cancellation system reduces this 50~Hz component to about
10\% of its initial value, Fig.~\ref{B_Komp}b.

\subsection{Lifetime of the D$_{5/2}$ state \label{lifetime}}
\begin{figure}[b]
\begin{minipage}{0.45\linewidth}
\begin{center}
\epsfig{file=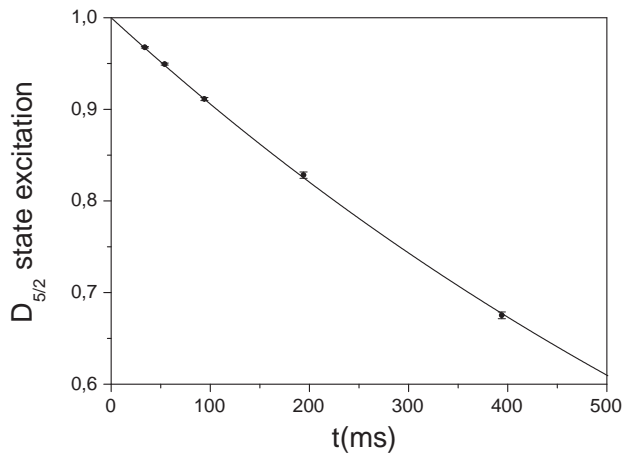, height=6cm}
\end{center}
\end{minipage}
\begin{minipage}{0.55\linewidth}
\caption{\label{decay} Lifetime measurement of the D$_{5/2}$
state. The collection of data points correspond to a total number
of $1.5\cdot10^5$ experiments. The errors bars indicate the
statistical error. The total data acquisition time was 6.5~h. We
find $1011\,(6)$~ms, where the error accounts for the statistical
error only.}
\end{minipage}
\end{figure}

The natural lifetime of about $\tau \simeq$ 1~s of the D$_{5/2}$
state sets a coherence limit for the single qubit. In order to
test the experiment, a single ion is excited to the D$_{5/2}$
state. We record the probability to find the ion in still in the
D$_{5/2}$ state after a waiting time $t$, see Fig.~\ref{decay}.
The resulting value of 1011~ms is slightly below the reported
value, see ref.~\cite{BARTON00}~Fig.~1. As discussed in
\cite{BARTON00,BLOCK99}, the D$_{5/2}$ decay is increased by
optical pumping to the ground state via the P$_{3/2}$ state with
residual light near 854~nm (see Fig.~1), thus explaining the
observation of a shorter lifetime
\cite{BARTON00,GUDJONS96,KNOOP95}.

In the experiment, the laser light near 854~nm was switched off
using a acousto-optical modulator in double pass configuration
with a measured attenuation of $\simeq 2\cdot10^{-4}$. Since
residual light near 854~nm still affects the observed lifetime, we
reduced the light power before the acousto-optical modulator for
the measurement in Fig.~\ref{decay}. As a second reason for a
reduction of $\tau$ the spontaneously emitted broadband background
radiation from the diode laser near 866~nm could have contributed
\cite{BARTON00}.

We do not claim to add another value to the literature for the
D$_{5/2}$ state lifetime, since this would require a profound
investigation of all systematic errors. However, from the above
measurements it becomes evident that any light resonant with
dipole transitions has to be suppressed. Note, that the coherence
of a superposition of the $|S\rangle$ and $|D\rangle$ states would
be affected also by an insufficient switching of the laser field
near 397~nm. For the light near 397~nm we use two acousto-optical
switches and an optical single mode fiber to avoid scattered light
(total isolation $\simeq 2\cdot10^{-6}$).

\section{Coherence of the motional qubit state}

To test the coherence of the motional qubit state, we  excite the "motional
qubit", represented by the vibrational states $\{|0\rangle,|1\rangle\}$. For
this , the laser frequency detuning is set to $\omega_{ax}$, the motional
sideband frequency, from the S-D resonance.

\subsection{Heating rate \label{heatingrate}}

If the ion's vibrational state increases without laser light
interaction, we denote this as a heating rate. Systematic studies
of heating rates have been performed in a three-dimensional Paul
trap \cite{ROOS99} and in linear traps \cite{ROHDE01} for single
ions and two-ion crystals. To measure the heating rate, the ion(s)
are first cooled to the vibrational ground state. Then the system
is left alone for a certain delay time $t$  to interact with the
environment, i.e. with the surrounding electrodes and any possible
perturbations acting on the motion of the ions. Finally the
resulting vibrational state is analyzed.

\begin{figure}[t]
\begin{center}
\epsfig{file=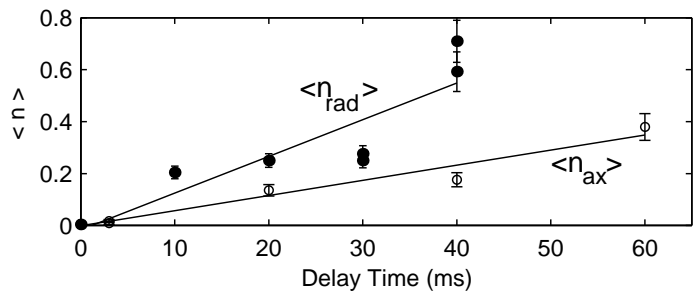, height=4cm}
\end{center}
\caption{\label{heating} Heating rate measurements for the axial
and radial vibrational modes at 4~MHz and 2~MHz, respectively.
Heating rates are 1 phonon in 190~ms for the axial and 1 phonon in
70~ms for the radial mode. Construction of the three-dimensional
trap: Ring electrode made of 0.2~mm molybdenum wire with inner
diameter of 1.4~mm \cite{ROOS99}.}
\end{figure}

For all traps we find that on the average it takes about 100~ms to
pick up a single phonon. We measure the increasing phonon number
by observing the Rabi--flopping signal on the blue sideband. The
results of such measurements,  axial phonon number $\langle n_{ax}
\rangle$ is as a function of the delay time $t$ are shown in
Fig.~\ref{heating}. Here, the data obtained for a single ion in
the three-dimensional Paul trap, result in a heating rate of
$d\langle n \rangle /dt = 0.0053$~ms$^{-1}$ (i.e. 1 phonon in 190
ms) at the trap frequency of $\omega_{ax}/(2\pi)$ = 4~MHz. For the
radial y direction the heating rate is determined to be 1 phonon
in 70 ms at $\omega_{rad}/(2\pi)$ = 1.9~MHz.

While the determination of the heating rate corresponds to the
measurement of the $T_1$-time, the following section is dedicated
to a measurement of the $T_2$-time, quantifying the dephasing of
superposition states.

\subsection{Decoherence of motional superpositions}
To successfully perform two-ion gate operations it is necessary
that the motional superpositions do not decohere. Similar to the
electronic (internal) coherence, we test this with a Ramsey
experiment. Therefore we prepare a superposition of two motional
states with identical electronic state. To achieve this we apply a
$\pi/2$-pulse on the carrier to obtain $(|n=0,\textrm{S}\rangle +
|n=0,\textrm{D}\rangle)/\sqrt{2}$. Secondly we apply a $\pi$-pulse
on the blue sideband. This pulse moves the
$|n=0,\textrm{S}\rangle$-part of the superposition into the
$|n=1,\textrm{D}\rangle$-state and leaves the
$|n=0,\textrm{D}\rangle$-part unaffected and we obtain the desired
superposition $(|n=1,\textrm{D}\rangle +
|n=0,\textrm{D}\rangle)/\sqrt{2})$. We now wait for a time $T$
before we apply the two laser pulses in the time reversed order.
If no dephasing has occured during $T$, this inverse pulse
sequence leads to the initial state $|n=0,\textrm{S}\rangle$, or
$|n=0,\textrm{D}\rangle$, depending on the phase of the last laser
pulse. Motional decoherence, however, would scramble the phase
relation between the $|n=0,\textrm{D}\rangle$ and
$|n=1,\textrm{D}\rangle$ state and thus reduce the contrast.

\begin{figure}[t]
\begin{minipage}{0.6\linewidth}
\begin{center}
\epsfig{file=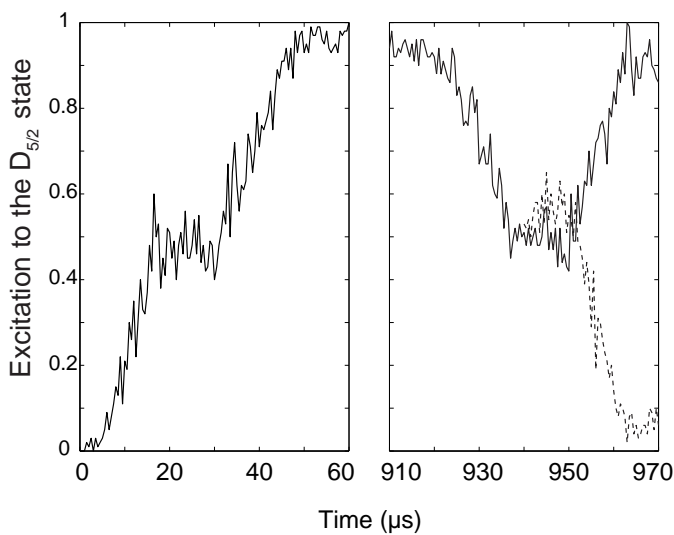, height=7cm}
\end{center}
\end{minipage}
\begin{minipage}{0.4\linewidth}
\caption{\label{motdecoherence} Decoherence of a superposition
state of $|n=0\rangle$ and $|n=1\rangle$, single ion in the linear
trap, for details see text.}
\end{minipage}
\end{figure}

Fig.~\ref{motdecoherence} shows the evolution of the D$_{5/2}$
state population of the excited state $P_D$ as a function of the
time. For this measurement we repeated the experimental cycle 100
times for each data point, however, we cut off the pulse sequence
at the time $t$ indicated on the $time$-axis. The first pulse is
complete at $t=20~\mu$s, the second one takes place from
$t=30~\mu$s to $60~\mu$s. All laser interaction is interrupted for
the interval $T$. Then the pulses are repeated in reversed order.
Depending on the relative phases of the first and the fourth pulse
we detect the atom in either the $|D\rangle$ (solid curve) or the
$|S\rangle$ state (dashed). For a waiting time $T=850~\mu$s we
measure a contrast of 80\%, mainly limited by the electronic
decoherence which takes place within each pair of pulses.
Observing the contrast for various interval times $T$ yields a
coherence time of approximately 100~ms for the motional
superposition state, about equal to the motional heating time.

\section{Future \label{future}}
With a systematic investigation of decoherence sources,
improvements of a future setup can be devised. We have found for
the qubit based on the S$_{1/2}$ to D$_{5/2}$ transition in
$^{40}{\rm Ca}^+$ that magnetic field fluctuations and laser
frequency fluctuations are the major limitations of the present
setup. The electronic qubit coherence time is about 1~ms. We
measure a coherence time of the motional qubit of about 100~ms.

Major progress is expected from switching from $^{40}{\rm Ca}^+$
to $^{43}{\rm Ca}^+$, an isotope with nuclear spin 7/2 which
posesses (i)~a hyperfine-split ground state and thus long-lived
states which may be coupled by Raman transitions, and
(ii)~magnetic-field insensitive Zeeman substates which are ideally
suited as qubit levels. Both these properties contribute to
potentially much longer coherence times of  $^{43}{\rm Ca}^+$,
since ambient magnetic fields affect the qubit phase only through
the much smaller quadratic Zeeman effect, since spontaneous
emission is practically absent, and since the laser phase on the
Raman transition is controlled by radio frequency technology and
is therefore less susceptible to laser noise. iii)~By
appropriately choosing the directions of the Raman beams with
respect to the trap axes, the Lamb-Dicke factor can be optimized
such that only the axial, but no radial vibrational modes are
excited. Therefore, radial spectator modes do not affect the gate
operation at all. In addition, the use of Raman transition speeds
up quantum gate operations since the two Raman photons transfer a
higher recoil to that mode used for the quantum gate
\cite{STEANE00}.

\begin{figure}[]
\begin{minipage}{0.45\linewidth}
\begin{center}
\epsfig{file=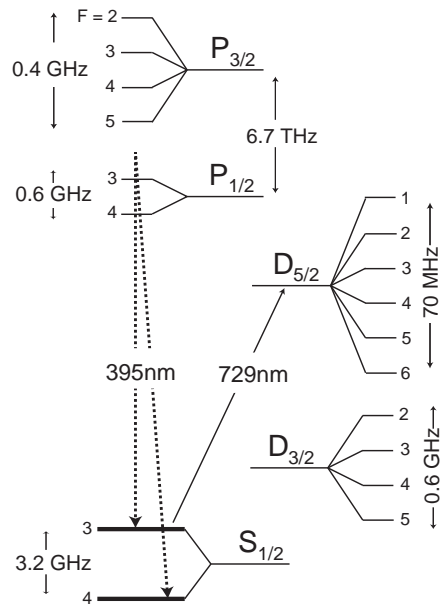, height=8cm}
\end{center}
\end{minipage}
\begin{minipage}{0.55\linewidth}
\caption{\label{ca43} Level scheme of $^{43}{\rm Ca}^+$. Qubit information
could be stored in the hyperfine ground states $|S_{1/2}, F=3, m=0\rangle$
and $|S_{1/2}, F=4, m=0\rangle$. A Raman transition near 396~nm can be used
for qubit manipulation. Prior to the state selective electron shelving, we
plan to transfer the $F=3$ state to the D$_{5/2}$ with a $\pi$-pulse at
729~nm.}
\end{minipage}
\end{figure}

Therefore, we plan to use $^{43}{\rm Ca}^+$ for quantum computing
\cite{STEANE97} and to encode the qubit in hyperfine ground states
$|S_{1/2}, F=3, m=0\rangle$ and $|S_{1/2}, F=4, m=0\rangle$.
Qubits can be manipulated on the Raman transition, as is indicated
in Fig.~\ref{raman_schema}. Doppler cooling may be performed
similarly as in $^{40}{\rm Ca}^+$ on the dipole transitions
S$_{1/2}$--P$_{1/2}$. For reaching the vibrational ground state,
either sideband cooling either on the S$_{1/2}$--D$_{5/2}$
transition, or on the Raman transition between both hyperfine
levels can be employed. For the quantum state detection we will
transfer one of the qubit levels, e.g. $|S_{1/2}, F=3,
m=0\rangle$, to the D$_{5/2}$ state by a resonant carrier $\pi$
pulse, similar to the scheme presented in Fig.~\ref{raman_schema}.
Eventually, the final state analysis is performed similarly as in
the case of $^{40}{\rm Ca}^+$ by detecting the fluorescence
emitted when the ion is illuminated by resonant radiation near
397~nm and 866~nm.

With strings of $^{43}{\rm Ca}^+$ we may be able to demonstrate,
at the proof-of-concept level, an ion trap quantum computer, using
the best features of today's existing experiments thus merging
advantages of experiments on $^{40}{\rm Ca}^+$ ions with
individual addressing and optical qubits, and on $^{9}{\rm Be}^+$
ions with hyperfine encoding of qubits \cite{WEBBOULDER}.

\subsection{Acknowledgments}
We gratefully acknowledge support by the European Commission
(QSTRUCT and QI networks, ERB-FMRX-CT96-0077 and -0087, QUEST
network, HPRN-CT-2000-00121, QUBITS network, IST-1999-13021), by
the Austrian Science Fund (FWF, SFB15), and by the Institut für
Quanteninformation GmbH.

\hspace{0.5cm}

\end{document}